\documentclass{article}
\usepackage{spconf,amsmath,graphicx,amssymb,amsfonts}

\usepackage{enumitem}
\usepackage{cite}
\usepackage{amsmath,amssymb,amsfonts}
\usepackage{algorithmic}
\usepackage{graphicx}
\usepackage{adjustbox}
\usepackage{multirow}
\usepackage{textcomp}
\usepackage{color,soul}
\usepackage{xcolor}
\usepackage{hyperref}
\setlist{nosep, leftmargin=14pt}
\usepackage{enumitem}
\setlist{nosep, leftmargin=14pt}

\usepackage{mwe} 

\title{SwinVFTR: A Novel Volumetric Feature-learning Transformer for 3D OCT Fluid Segmentation}
\name{Author(s) Name(s)\thanks{Some author footnote.}}
\name{Khondker Fariha Hossain$^{\dagger}$$^{\star}$\thanks{$^{\star}$ Equal Contribution}, Sharif Amit Kamran$^{\ddagger}$$^{\star}$, Alireza Tavakkoli$^{\dagger}$, George Bebis$^{\dagger}$, Sal Baker$^{\mathsection}$}
\address{$^{\dagger}$ Department of Computer Science \& Engineering, University of Nevada, Reno, NV, USA\\
$^{\mathsection}$ School of Medicine, University of Nevada, Reno, NV, USA\\
$^{\ddagger}$ Johnson \& Johnson Innovative Medicine}
\begin{document}
%
\maketitle
\begin{abstract}
Accurately segmenting fluid in 3D optical coherence tomography (OCT) images is critical for detecting eye diseases but remains challenging. Traditional autoencoder-based methods struggle with resolution loss and information recovery. While transformer-based models improve segmentation, they aren’t optimized for 3D OCT volumes, which vary by vendor and extraction technique. To address this, we propose SwinVFTR, a transformer architecture for precise fluid segmentation in 3D OCT images. SwinVFTR employs channel-wise volumetric sampling and a shifted window transformer block to improve fluid localization. Moreover, a novel volumetric attention block enhances spatial and depth-wise attention. Trained using multi-class dice loss, SwinVFTR outperforms existing models on Spectralis, Cirrus, and Topcon OCT datasets, achieving mean dice scores of 0.72, 0.59, and 0.68, respectively, along with superior performance in mean intersection-over-union (IOU) and structural similarity (SSIM) metrics.
\end{abstract}
\begin{keywords}
Fluid Segmentation, Optical Coherence Tomography, Swin Transformer, OCT Segmentation
\end{keywords}
\section{Introduction}
\label{sec:intro}

Fluid buildup, or macular edema in retinal layers, is a common reason for blindness and retinal degeneration. Possible factors include Drusen, Choroidal neovascularization (CNV), Age-related macular degeneration (AMD), and Diabetic retinopathy (DR) \cite{tranos2004macular}. Age-related macular degeneration causes irreversible blindness in approximately 8.7\% of people worldwide and is a leading cause of vision loss.  Similarly, Diabetic retinopathy affects one-third of every diabetic patient \cite{ting2016diabetic}, which is 2.8\% of the world's population and the second leading cause of blindness. As a result, early diagnosis, localization, and segmentation of retinal layer fluid accumulation can help with effective treatments. Optical coherence tomography (OCT) is a non-invasive retinal imaging method  that yields 3D volumetric cross-sectional images for viewing the morphology of retinal layers and underlying pathologies. Although the image is extracted through this approach, the differential diagnosis and fluid localization are supervised by an expert ophthalmologist. 

Manual annotation and segmentation of sub-retinal fluid are time-consuming, error-prone, and tedious. To address this, traditional image-processing and machine learning \cite{chiu2015kernel} techniques were introduced but relied heavily on handcrafted features, struggled with spatial and depth features, and lacked generalization ability. With deep learning, automated segmentation in medical imaging gained traction due to its high accuracy in pixel-wise segmentation and volumetric data. 2D U-Net-like auto-encoder models \cite{he2022intra} have been widely used for retinal fluid segmentation, showing good results in multi-layer segmentation. However, they struggle with fine fluid boundaries and detecting small deposits. Vision transformer-based models have recently improved small fluid segmentation by using multi-headed \cite{wang2022tiny} or shifted-window \cite{philippi2023vision} attention to capture local context. However, these models operate on 2D slices and lack depth information. While early 3D U-Net models showed potential, progress has stagnated. We propose an approach to overcome these limitations:

\begin{figure*}[!htp]
    \centering
    \includegraphics[width=1\linewidth]{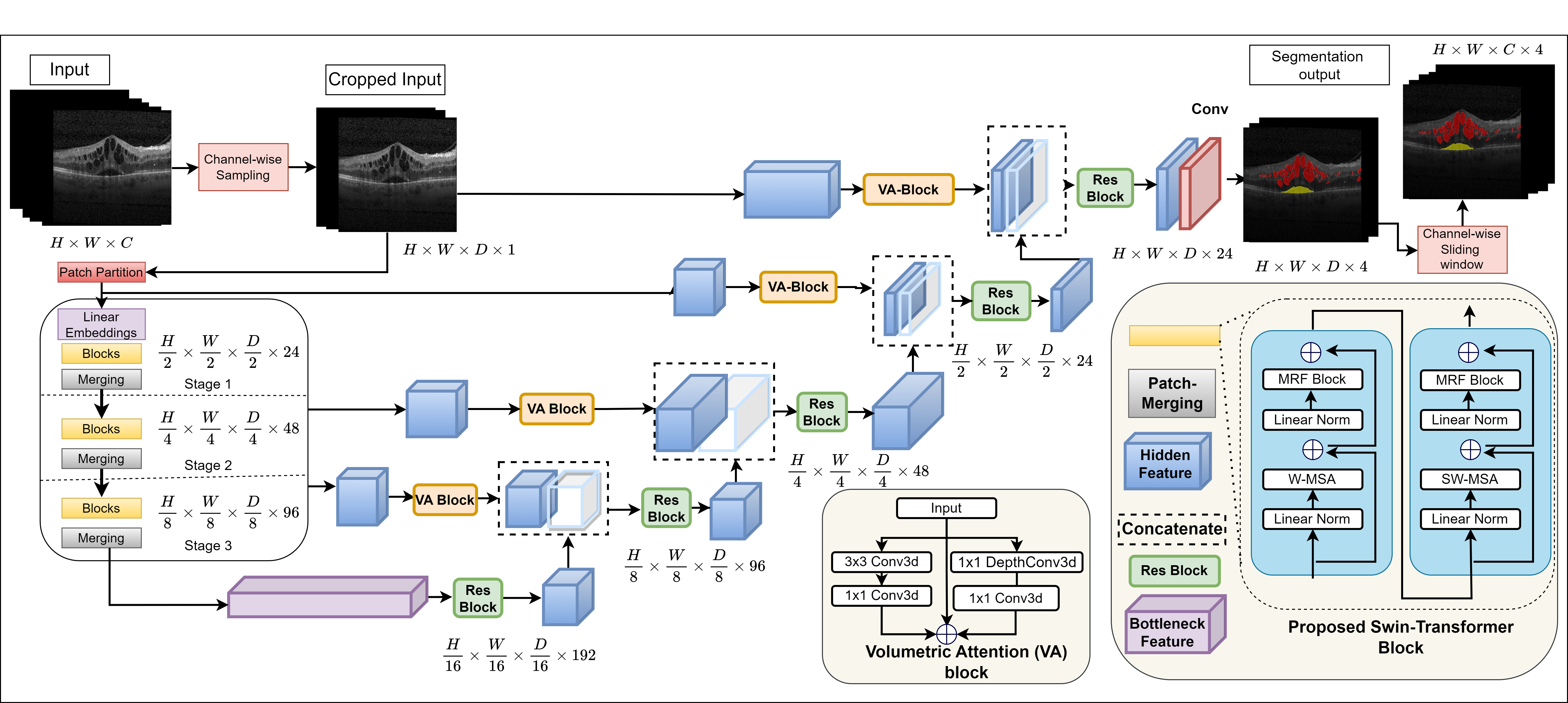}
    \caption{Proposed SwinVFTR architecture which takes 3D OCT volume input with channel-wise sampling technique and outputs a 3D segmentation map of the fluid accumulation. The SwinVFTR encoder incorporates a new swin-transformer block consisting of Shifted window attention, Multi-headed attention and Multi-Receptive Field (MRF) sub-block with both convolution and dilated convolution layers. The encoder features are sequentially added with the decoder using a skip-connection consisting of a volumetric attention (VA) block.}
    \label{fig1}
\end{figure*}
\begin{itemize}
    \item[$\bullet$] A novel architecture termed Swin Volumetric Feature-learning Transformer (SwinVFTR) that utilizes a swin-transformer as an encoder and joins it to a 3D convolution-based decoder at distinct resolutions via novel volumetric spatial and depth attention block. Moreover, we modify the swin-transformer block with a Multi-receptive field residual block instead of MLP.
    \item[$\bullet$] Our model employs a channel-wise overlapped sampling technique to crop OCT volumes only in the depth axis while retaining spatial information. We minimize data loss due to resizing 3D volumes by replacing it with channel-wise sampling for inference.
    \item[$\bullet$] To validate our work, we compare four different 3D convolution and transformer-based architectures for medical image segmentation on three vendor-specific OCT datasets: Spectralis, Cirrus, and Topcon \cite{bogunovic2019retouch}. From Fig.~\ref{fig2}, it is apparent that our architecture segments retinal fluid with high dice-score and mean-IOU.
\end{itemize}

\section{Methodology}
\label{sec:methodology}

\subsection{Channel-wise Volumetric Sampling}
Sampling OCT B-scans at different depths can affect the outcome of recognizing retinal disease pathology for accurate diagnosis. Although U-Net-like architectures are flexible in handling OCT volumes of different depths, current transformer-based architecture cannot take OCTs with smaller depths. For example, UNETR \cite{hatamizadeh2022unetr} and Swin-UNETR \cite{tang2022self}, two state-of-the-art models for medical image segmentation, utilize patch-merging layer to downsample $\times 32$. As a result, any OCT with less than 64 B-scans cannot be used out-of-the-box for these models. Since we are working on diversified OCT volumes with B-scans of 49 to 128, utilizing volumetric cropping would be ideal. However, we want to retain the spatial information while sampling a section of the original B-scans. So we introduce a channel-wise sampling technique that samples a cropped image with dimension of $\mathbb{R}^{H \times W \times D}$,  from an image with $\mathbb{R}^{H \times W \times C}$ dimensions, where $D < C$ and $D=32$. We also utilize one less swin-transformer and patch-merging block to make our downsampling $\times 16$. While producing the output, we do channel-wise overlapped volume stitching (25\% overlap), which is given in Fig.~\ref{fig1}.

\subsection{Proposed Swin-Transformer Block}
Regular window-based multi-head self-attention (W-MSA), which was incorporated in Vision Transformer (ViT) \cite{dosovitskiyimage}, employs a single low-resolution window for constructing a global feature map and has quadratic computation complexity. Contrastly, the Swin Transformer architecture proposed in \cite{liu2021swin} integrates shifted windows multi-head self-attention (SW-MSA), which builds hierarchical local feature maps and has linear computation complexity. Recently, Swin-UNETR \cite{tang2022self} adopted this swin-transformer block without making any fundamental changes and achieved state-of-the-art dice scores in different 3D medical image segmentation tasks. One of the most significant drawbacks of these blocks is using a Multi-layer perceptron block (MLP) after the post-normalization layer. MLP utilizes two linear (dense) layers, which are computationally more expensive than a 1D convolution with a small kernel. For example, a linear embedding output from a swin-transformer layer having dimension $X$ (where $X \in \mathbb{R}^{H\times W\times D}$), and input channel, $C_{in}$ and output channel, $C_{out}$ will have a total number of parameters, $\mathbb{R}^{X\times C_{in} \times C_{out}}$. In contrast, a 1D Conv with kernel size, $k$ with the same input and output will have less number of parameters, $\mathbb{R}^{k\times C_{in} \times C_{out}}$. Here, we did not consider any parameters for bias and the value of $k=\{1,3\}$. On the other hand, using 1D convolution will drastically affect performance, given that small receptive fields only account for local features, not global ones. Hence, we employ a multi-branch residual block with vanilla and dilated convolution termed Multi-receptive field Block (MRF) to address this. So for subsequent layers, $l$ and $l+1$, the proposed swin-transformer block can be defined as Eq.~\ref{eq1}.

\begin{equation}
    \begin{split}
    &d^{l} = W{\text -}MSA(\psi(d^{l-1})) + d^{l-1}\\
    &d^{l} = MRF(\psi(d^{l})) + d^{l} \\
    &d^{l+1} = SW{\text -}MSA(\psi(d^{l})) + d^{l}\\
    &d^{l+1} = MRF(\psi(d^{l+1})) + d^{l+1}
    \end{split}
    \label{eq1}
\end{equation}

In Eq.~\ref{eq1}, we visualize the first sub-block of the swin-transformer consisting of LayerNorm ($\psi$) layer, multi-head self-attention module (W-MSA), residual connection (+), and Multi-receptive field block (MRF). Similarly, the second sub-block of the swin-transformer consisting of LayerNorm ($\psi$) layer, shifted window multi-head self-attention module (SW-MSA), residual skip-connection (+), and Multi-receptive field block (MRF). Moreover, $l$ signifies the layer number, and $d$ is the feature-map. The MRF block can be further elaborated in Eq.~\ref{eq2}.

\begin{equation}
    \begin{split}
    &x^1 = \delta(Conv(x_{in}))\\
    &x^2 = \delta(Depthwise\_Conv(x^1)) \\
    &x^3 = \delta(Dilated\_Conv(x_{in})) \\
    &x_{out} =  \delta(Conv(x^1 + x^2 + x^3))
    \end{split}
    \label{eq2}
\end{equation}

In Eq.~\ref{eq2}, we first use convolution with kernel size, $k=1$, and stride, $s=1$ to extract local features with a small receptive field. Then, we insert the output of these layers into a depth-wise convolution layer ($k=1$, $s=1$). In a parallel branch, we use a dilated convolution ($k=3$, $s=1$) with a dilation rate of $d=2$ to extract features with a larger receptive field. Finally, we add all these outputs from these three convolution layers and then apply a convolution ($k=1$, $s=1$) to get the final result. Here, $\delta$ signifies the GELU activation, which is applied to all convolution layers.

\subsection{Encoder}
We transform the OCT volumetric images before the encoder can take the input with dimensions $\mathbb{R}^{H \times W \times D}$. We employ a patch partition step to create a sequence of 3D tokens with a dimension of $\mathbb{R}^{\frac{H}{P} \times \frac{W}{P} \times \frac{D}{P}}$ and these features are then projected to an embedding space with dimension $C$. Specifically, our encoder has a non-overlapping patch with a size of $2 \times 2 \times 2$ and a feature dimension of $2 \times 2 \times 2 \times 1  = 8$ by considering one channel of the OCT. We assign the embedding space size, C=24, in our encoder. So the feature output of the patch-parition layer is $\mathbb{R}^{\frac{H}{2} \times \frac{W}{2} \times \frac{D}{2} \times 24}$ .  Likewise, each encoder stage downsamples the features by utilizing two swin-transformer blocks followed by a patch-merging block. So, the  features size changes from $\mathbb{R}^{\frac{H}{2} \times \frac{W}{2} \times \frac{D}{2} \times C}$ to $\mathbb{R}^{\frac{H}{4} \times \frac{W}{4} \times \frac{D}{4} \times 2C}$, from $\mathbb{R}^{\frac{H}{4} \times \frac{W}{4} \times \frac{D}{4} \times 2C}$ to $\mathbb{R}^{\frac{H}{8} \times \frac{W}{8} \times \frac{D}{8} \times 4C}$, and from $\mathbb{R}^{\frac{H}{8} \times \frac{W}{8} \times \frac{D}{8} \times 4C}$ to $\mathbb{R}^{\frac{H}{16} \times \frac{W}{16} \times \frac{D}{16} \times 8C}$, successively. We incorporate two swin-transformer blocks after the last patch-merging layer to finalize the encoder.

\subsection{Volumetric Attention Block}
In 3D UNet-like architectures \cite{hatamizadeh2022unetr}, skip connections concatenate the encoder and decoder features to retain loss of information. However, to make these features more robust, Swin-UNETR incorporated residual attention block with two convolution layers \cite{oktayattention,kerfoot2019left}. The problem with this approach is that it utilizes regular convolution, which only applies attention spatially and ignores any channel-wise attention. To alleviate this, we propose a volumetric attention (VA) block of separate branches. In the first branch, we have a $3 \times 3 \times 3$ followed by a $1 \times 1 \times 1$ convolution for attention in the spatial dimension ($H\times W$). In the following branch, we have a $1 \times 1 \times 1 $ depth-wise convolution followed by a $1 \times 1 \times 1$ point-wise convolution for depth and channel-wise attention ($D\times C$). In the final branch, we have an identity function that copies the input features. Consequently, we add these features to generate our last output feature.

\subsection{Decoder}
Similar to our encoder, we design a symmetric decoder composed of multiple transposed convolution blocks and a volumetric concatenation layer between each stage of the encoder and decoder features. At each stage n ($n \in {1,2,3}$) in the encoder and bottleneck ($n=4$), the volumetric feature representations are reshaped to $\mathbb{R}^{\frac{H}{2^n} \times \frac{H}{2^n} \times \frac{H}{2^n}}$ and inserted into a residual convolution block with two $3 \times 3 \times 3$ convolution followed by instance normalization layer. Each decoder's feature maps are doubled in size using a transposed convolution layer. Moreover, each encoder's skip feature maps through the VA blocks are concatenated with the outputs of the previous decoder. Finally, a residual convolution block is applied to the feature with two $3 \times 3 \times 3$ convolutions followed by an instance normalization layer. The final segmentation output is generated using a $1 \times 1 \times 1$  convolutional layer and a softmax activation function.

\begin{figure}[tp!]
    \centering
    \includegraphics[height=7cm,width=1\linewidth]{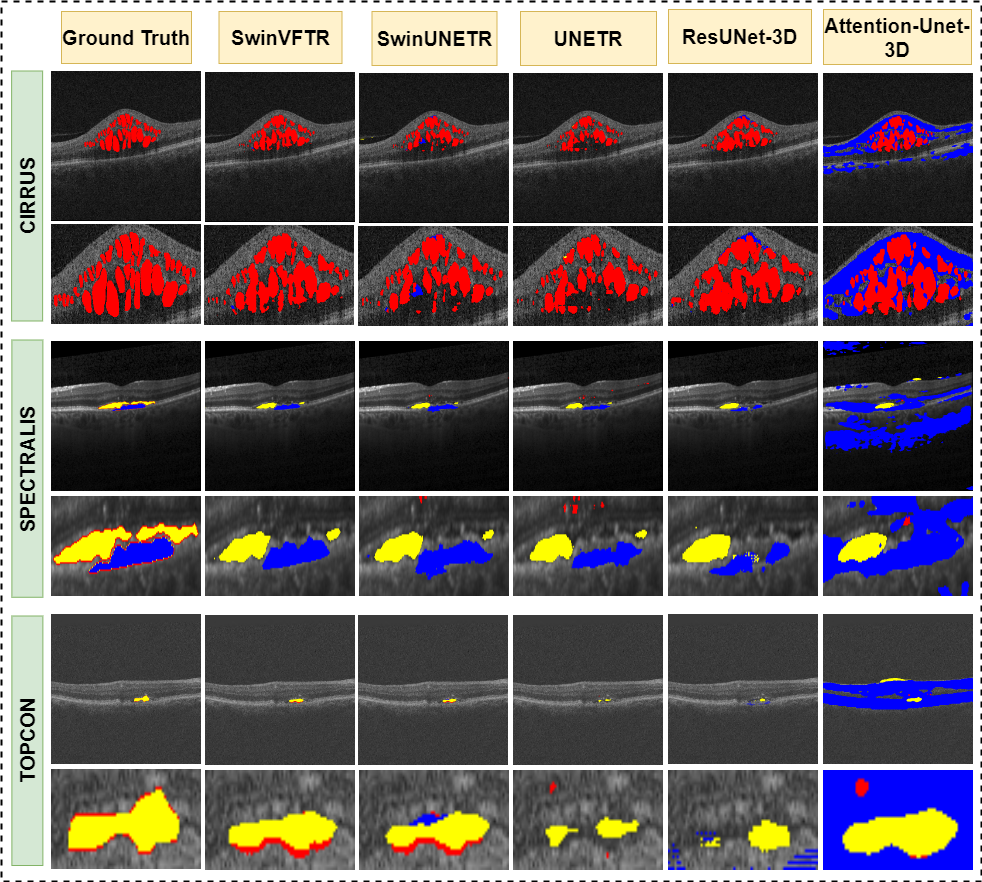}
    \caption{SwinVFTR segments fluid with better precision than other 3D CNN and Transformer architectures. The row contains Cirrus, Spectralis and Topcon data-sets. Whereas the column contains ground-truths and segmentation maps for SwinVFTR, SwinUNETR, UNETR, ResUNet-3D and Attention-UNet-3D. Here, IRF, SRF, and PED fluids are colored as {\color{red} Red}, {\color{yellow} Yellow} and {\color{blue} Blue}.}
    \label{fig2}
\end{figure}

\section{Experiments}
\label{sec:experiment}

\subsection{Dataset and Preprocessing}
For benchmarking, we use the RETOUCH public dataset \cite{bogunovic2019retouch}, which contains three image sets from three unique vendor devices and, in total, has 70 volumes. Out of this, 24 volumes were obtained with Cirrus (Zeiss), 24 volumes with Spectralis (Heidelberg), and 22 volumes with T-1000 and T-2000 (Topcon) devices. The numbers of B-scans (volume depths) were 128, 49, and 128, with resolutions of 512×1024, 512×496, and 512×885, respectively, for each of these vendor devices. We only resize Cirrus and Topcon volumes to $512\times 512$ resolution. The volume contained three different fluids such as intra-retinal fluid (IRF), sub-retinal fluid (SRF), and pigment epithelial detachment (PED), . We separate the original image sets into training-validation and test set. So for Cirrus and Spectralis, we had 19 training-validation and 5 test volumes, whereas, for Topcon, we had 18 training-validation and 4 test volumes. We utilize 5-fold cross-validation to find the model with the highest dice score. For image transformations, we apply Random Intensitity Shift ( +/- 10 with 50\% probability) and Random Channel-wise volumetric cropping. 
\subsection{Hyper-parameter Initialization}
We used Adam optimizer, with learning rate $\alpha=0.0001$, $\beta_1=0.9$ and $\beta_2=0.999$. We train with mini-batches with batch size, $b=1$ for 600 epochs using PyTorch and MONAI library.It took between 8-12 hours to train our model on NVIDIA A30 GPU. The inference time is $0.5$ second per volume. The code repository is provided in this \href{https://github.com/SharifAmit/Swin-VFTR}{link}. 

\begin{table}[!tp]
\centering
\caption{Quantitative comparison on Spectralis,  Cirrus, \&  Topcon \cite{bogunovic2019retouch}.}
\begin{adjustbox}{width=1\linewidth}
\begin{tabular}{|c|c|c|c|c|c|c|c|c|c|}
\hline
\multirow{2}{*}{Dataset} & \multirow{2}{*}{Method} & \multirow{2}{*}{Year}  & \multirow{2}{*}{SSIM}  & \multirow{2}{*}{MIOU}  & \multicolumn{3}{c|}{Dice Score}  &  \multicolumn{2}{c|}{Mean Dice} \\ \cline{6-10} 
& & & & & IRF & SRF & PED & w/o BG & w/ BG \\ \hline
\multirow{5}{*}{Spectralis}  & Attn-UNET3D \cite{oktayattention} & 2018 &  0.914 &   0.439 &  0.608 & 0.394 & 0.096 & 0.366 & 0.517 \\ 
& ResUNet-3D\cite{kerfoot2019left} & 2019 & 0.985 & 0.595 & 0.602 & 0.574 & 0.602 & 0.592  &  0.694 \\ 
& UNETR \cite{hatamizadeh2022unetr} & 2022 & 0.984 & 0.546 & 0.567 & 0.493 & 0.544 & 0.534 & 0.651 \\ 
& SwinUNETR \cite{tang2022self} & 2023 & 0.985 & 0.613 & 0.601 & 0.544  & 0.662 & 0.602 &  0.701     \\ 
& \textbf{SwinVFTR} & 2024 & \textbf{0.987} & \textbf{0.625} & \textbf{0.624} & \textbf{0.578} & \textbf{0.670} & \textbf{0.624} & \textbf{0.718}  \\\hline	
\multirow{5}{*}{Cirrus}& Attention-UNet-3D   \cite{oktayattention} & 2018 & 0.928 & 0.446 & 0.664 & 0.472 & 0.011 & 0.382 & 0.527\\
& ResUNet-3D \cite{kerfoot2019left}  & 2019 &  0.983 & 0.490 & 0.648 & \textbf{0.622} & 0.012 & 0.427 & 0.570 \\ 
& UNETR \cite{hatamizadeh2022unetr} & 2022 & 0.987 & 0.487 & 0.635 & 0.594 & 0.081 &  0.436 & 0.577     \\
& SwinUNETR \cite{tang2022self} & 2023 & 0.986 & 0.452 & 0.682 & 0.338 & 0.131 & 0.384 & 0.537\\
& \textbf{SwinVFTR} & 2024 & \textbf{0.988} & \textbf{0.492} & \textbf{0.691} & 0.507 & \textbf{0.146} & \textbf{0.448} & \textbf{0.587}  \\\hline
\multirow{5}{*}{Topcon}& Attention-UNet-3D \cite{oktayattention} & 2018 & 0.894 & 0.412 & 0.526 & 0.542 & 0.083 & 0.383 & 0.519\\
& ResUNet-3D \cite{kerfoot2019left}  & 2019  & 0.974 & 0.526 & 0.534 & 0.\textbf{648} & 0.419 & 0.534 & 0.649 \\ 
& UNETR \cite{hatamizadeh2022unetr} & 2022 & 0.979 & 0.495 & 0.592 & 0.416 & 0.427 & 0.478 &  0.607      \\
& SwinUNETR \cite{tang2022self} & 2023 & 0.980 & 0.483 & 0.583 & 0.451 & 0.331 & 0.455 & 0.590 \\
& \textbf{SwinVFTR} & 2024 & \textbf{0.981} & \textbf{0.553} & \textbf{0.638} & 0.523 & \textbf{0.548} & \textbf{0.571} & \textbf{0.678}  \\\hline	
\end{tabular}
\label{table1}
\end{adjustbox}
\end{table}

\subsection{Quantitative Evaluation}
We compared our architecture with some best-performing 3D CNN and Transformer architectures, including ResUNet-3D \cite{kerfoot2019left}, AttentionUNet-3D \cite{oktayattention}, UNETR \cite{hatamizadeh2022unetr} and SwinUNETR \cite{tang2022self} as illustrated in Fig.~\ref{fig2}. We trained and evaluated all four architectures using their publicly available source code on the three datasets. SwinUNETR utilizes a swin-transformer encoder as a backbone and a step-wise decoder with transposed convolution and residual blocks to upsample the features. In contrast, the UNETR employs a vision transformer with self-attention layers as encoders and deconvolution layers for upsampling. ResUnet-3D and Attention-UNet-3D are simple modifications of UNet 3D architectures, with the first using residual layers and the second incorporating attention layers. In Fig.~\ref{fig2}, we visualize segmentation results for intra-retinal fluid (IRF), sub-retinal fluid (SRF), and pigment epithelium detachments (PED). It is apparent from the figure that our model's prediction is more accurate than other transformer and CNN-based architectures, and the segmentation boundary is finer and less coarse than SwinUNETR and UNETR. Next, we quantitatively evaluate all five models using mean-intersection-over-union (mIOU), dice scores, and structural similarity index (SSIM) as shown in Table.~\ref{table1}. We also provide fluid-wise dice scores for IRF, SRF, and PED. Table.~\ref{table1} shows that our model's overall dice score, SSIM, and mIOU far exceed other architectures. Although our model's segmentation performance for Cirrus and Topcon is a little worse for SRF fluid against ResUNet-3D, the dice score PED is almost $10\times$ better for Cirrus and $1.2\times$ better for Topcon. We also calculate dice score with (w/ BG) and without background (w/o BG), as background contains the majority of the pixels, and it can skew the results with high false-positive rates. As the table shows, our model outperforms other architectures with a higher dice score for both with and without background.

\section{Conclusion}
\label{sec:conclusion}

In this paper,  we proposed a new 3D transformer-based fluid segmentation architecture called SwinVFTR. Combining our novel channel-wise sampling technique, incorporating volumetric attention, and using a multi-receptive field swin-transformer block, the architecture segments fluid volumes with high precision for three relevant metrics. We hope to extend this work to other ophthalmic modalities.

\section{ACKNOWLEDGEMENTS}
This work was partially supported by the National Institute of General Medical Sciences of the National Institutes of Health under grant number P30 GM145646 and by the National Science Foundation under grant number OAC 2201599 and grant number OIA 2148788..

\section{COMPLIANCE WITH ETHICAL STANDARDS}
This research study was conducted retrospectively using human subject data made available in open access by \cite{bogunovic2019retouch}. Ethical approval was not required, as confirmed by the license attached with the open-access data.

\bibliographystyle{IEEEbib}
\bibliography{refs}

\end{document}